\documentclass[12pt,preprint]{aastex}

\shorttitle{Hubble ACS mosaic of M82}
\shortauthors{Mutchler et al.}

\begin{document}


\title{{\it Hubble Space Telescope} ACS mosaic \\
    of the prototypical starburst galaxy M82}


\author{M. Mutchler, H. E. Bond, C. A. Christian, L. M.
Frattare, \\ F. Hamilton, W. Januszewski, Z. G. Levay, M. Mountain, K. S. Noll,
P. Royle}
\affil{Space Telescope Science Institute, 3700 San Martin Drive, Baltimore, MD 21218}
\email{mutchler@stsci.edu}

\author{J. S. Gallagher}
\affil{University of Wisconsin}

\and

\author{P. Puxley}
\affil{National Science Foundation}




\begin{abstract}

In March 2006, the Hubble Heritage Team obtained a large 4-filter (B, V, I, and
H\( \alpha  \)) 6-point mosaic dataset of the prototypical starburst galaxy NGC
3034 (\objectname{M 82}), with the Advanced Camera for Surveys (ACS) onboard the
{\it Hubble Space Telescope} (HST). The resulting color composite Heritage image
was released in April 2006, to celebrate Hubble's 16th anniversary.

Cycle 15
HST proposers were encouraged to submit General Observer and Archival Research
proposals to complement and/or analyze this unique dataset. Since our
\objectname{M 82} mosaics represent a significant investment of expert
processing beyond the standard archival products, we will also release our
drizzle-combined FITS data as a High-Level Science Product via the Multimission
Archive at STScI (MAST) in December 2006.

This paper documents the key aspects of the observing program and image
processing: calibration, image registration and combination (drizzling), and the
rejection of cosmic rays and detector artifacts. Our processed FITS mosaics and
related information can be downloaded from:

\url{http://archive.stsci.edu/prepds/m82/}

\end{abstract}



\section{Introduction}

Last year, the {\it Hubble} and {\it Spitzer} space telescopes made
complementary optical and infrared mosaics, respectively, of The Whirlpool
Galaxy M51, see \citet{mut05,cal05}. This year, the starburst galaxy M82
received similar treatment, described by this paper and \citet{eng06},
respectively. Although we released the color composite image of the Hubble M82
mosaic in April 2006 (see press release URL below), the science data are
embargoed until December 2006.

\objectname{M82} hosts the nearest (D=3.6 Mpc) example of a giant starburst
galaxy, with many key processes, such as super star clusters, ISM interfaces,
and outflows, occurring on scales that benefit from Hubble's high angular
resolution for their study. 

\section{Observations}


Our observing program (HST proposal 10776) produced 96 individual exposures with
the ACS Wide Field Channel (WFC). The WFC is composed of two 4096x2048 pixel CCD
arrays butted together, with a small gap (equivalent to 50 pixels wide) between
them. The total WFC field-of view is about  200 square arcseconds. For each of
the four filters (Johnson B,V,I, and H\( \alpha  \)), four exposures were
obtained at six slightly overlapping pointings or {}``tiles{}'' in a 3x2 mosaic,
at a telescope orientation of 130 degrees. The exposure times
for each filter are summarized in Table 1.  

The four exposures within each tile
were dithered. A small sub-pixel dither (2.5 x 1.5 pixels) improves the
rejection of detector artifacts as well as cosmic rays, coupled with a larger
dither which spans the interchip gap (5x60 pixels), so the combined mosaic will
not have strips of missing data there. These patterns are fully described on the
ACS dither webpage:

\url{http://www.stsci.edu/hst/acs/proposing/dither/}

\section{Pipeline data calibrations}

The standard archival data was retrieved and automatically calibrated {\it
on-the-fly}  via the Multimission Archive at STScI (MAST), after the best
calibration reference files became  available (a few weeks after the
observations).  This standard ACS pipeline (CALACS) processing includes bias,
dark, and  flat-field corrections.  The calibration reference files also
propagate data quality flags which identify many types of detector artifacts,
most of which were excluded during the {\tt MultiDrizzle} processing described
below. These include bad CCD columns, hot pixels (and their CTE tails), warm
pixels, and saturated pixels. For further details of standard calibrations are
provided in the ACS Data Handbook, \citet{pav05}. The latest calibration
reference files, and definitions of data quality flags are available on the ACS
website: 

\url{http://www.stsci.edu/hst/acs/analysis/reference\_files}

As input to the drizzle-combination described below, we used the archival
flat-fielded images ({\tt \_flt.fits} or hereafter simply {\tt flt}). Since this
dataset is not associated, the pipeline currently generates single-image
drizzled output ({\tt \_drz.fits}) for each exposure. Although {\tt
MultiDrizzle} was added to the ACS pipeline in 2004, the HST ground system can
only combine associated datasets, and in any case, cannot assemble large
mosaics. In the standalone environment, however, {\tt MultiDrizzle} can be used
to combine any set of images, generating it's own association table ({\tt
\_asn.fits}) for all the input exposures. Note that this table also gets updated
with the shifts and rotations provided in the {\tt shiftfile}, to register the
mosaic (as described below). 

\section{Image registration}

Our image combination software uses the World Coordinate System (WCS)
information in the image headers to align the images. This works very well for
data taken within the same visit, or otherwise utilizing the same guide stars.
Such data are typically aligned to within at least 0.2 WFC pixels (0.01 arcsec).
But mosaic datasets involving shifts larger than about 100 arcseconds
necessarily involve different guide stars in different visits, and so guide star
positional errors render the WCS unreliable for adequate image registration.

We use stars in the images to register the frames, and since more stars are
detected in the I-band (F814W), we use those images to measure  relative
positions and derive delta-shifts (relative to the WCS). We arbitrarily define
the first POS2 (visit 21) I-band frame (rootname {\tt j9l021d5q}) to be the
origin of our reference frame (0,0,0). Since the POS2 exposures for each filter
executed in the same visit, they are well-aligned to each other. Therefore, the
filter-to-filter registration is adequate, and we can apply the I-band shifts to
the other filters with confidence. After the I-band shifts were applied to the
other filters, their alignment was spot-checked, and individual stars across all
four filters were verified to be well-registered.

We use {\tt tweakshifts} for the final refinement of the shifts.
But {\tt tweakshifts} only works within the regime where the images
to be registered are already aligned to within about 1 pixel, and with little
or no rotation involved. So the large shifts between the tiles must
be removed before we can run {\tt tweakshifts}.

\subsection{Intertile shifts}

The data for each tile were obtained in different visits, so the tiles can
therefore be misaligned on the order of 10 pixels, and rotated with respect to
each other due to the nominal roll of the spacecraft. Fewer stars are available
to measure these ``intertile'' shifts, since they must lie in the narrow tile
overlaps. 

These initial intertile shifts were measured using only the first of the four
frames within each tile. {\tt MultiDrizzle} was run only through the   {\tt
driz\_separate} step, to generate the single-drizzled (distortion corrected {\tt
\_single\_sci.fits}) versions of these images in the undistorted mosaic output
space. We set {\tt {clean=no}} to save the {\tt \_single\_sci.fits} files, and
avoid altering the input {\tt flt} images by setting {\tt skysub=no} (no sky
subtraction). Suitable registration objects (preferably stars, but sometimes
clusters with a sharp peak) were visually selected: they must appear in multiple
frames, be free of obvious cosmic ray contamination, and not be saturated and
bleeding. With {\tt imexam}, the x,y positions of 10-20 objects were measured in
the tile overlap regions. The POS1, POS3, and POS5 tiles have long overlaps with
the POS2 reference tile, so objects were chosen which span the full length of
each overlap.  The POS1 and POS6 tiles have very small overlaps with POS2, so
their shifts were measured iteratively, i.e. after the corrective shifts and
rotations had been applied to POS1, POS3, and POS5. Then stars in their much
larger overlaps with POS1 and POS6 could also be used to measure shifts, with
greater leverage.

The measured star positions were used as input to the {\tt geomap} task, which
can solve for shifts and rotations. The corrections for the six primary
intertile frames were also applied to the other three frames in their respective
tiles, thereby internally registering the entire mosaic. The shifts and
rotations were assembled into a {\tt shiftfile}, to be provided as input to {\tt
MultiDrizzle}.

\subsection{Intratile shifts}

To generate the files needed for refining the shift measurement with   {\tt
tweakshifts}, {\tt MultiDrizzle} was again run only through the    {\tt
driz\_separate} step, applying the intertile delta-shifts derived above. Then
{\tt tweakshifts} was run for overlapping sets of I-band {\tt
\_single\_sci.fits} images, this time using all the images in each tile (not
just the first frame), such that the refinement includes any small ``intratile''
shifts. With a tolerance of 0.7 pixels and a 6 sigma threshold, {\tt
tweakshifts} typically detected on the order of 10,000 objects (most of which
are cosmic rays and other artifacts) and matched a few percent of those objects
within each tile. The rotations measured initially (as a check) were
insignificant, so we solved only for x,y shifts. These intratile refinements
were then summed with the initial intertile shifts, to create the shiftfiles
provided to {\tt MultiDrizzle} for the final image combinations.

\section{Masking artifacts}

In addition to the detector artifacts automatically flagged by the calibration
pipeline, we can manually mask observational artifacts that we'd similarly like
to exclude from our data, and which are too diffuse or large to get rejected
along with the cosmic rays. The variety of artifacts which might be seen are
described on the ACS anomalies webpage:

\url{http://www.stsci.edu/hst/acs/performance/anomalies}

First, we note two significant artifacts which we do not attempt to mask or
correct. This dataset was afflicted by some filter reflections (or ``optical
ghosts'') from several bright foreground stars, which appear as figure-8 or
doughnut features (see Figure 1). But since the ghosts overlap each other,
leaving no good data in any of the frames being combined, we chose not to mask
them.  Some amplifier crosstalk is also evident in the mosaics, H\( \alpha  \)
POS3 in particular, where both the bright star and galaxy create negative
mirror-image imprints in the opposite amplifier quadrant. This effect is not
significant, and we do not attempt to mask or correct it. 

Some satellite trails were masked and excluded as follows. The endpoints and
width of each satellite trail were measured in the undistorted {\tt
\_single\_sci.fits} images. Then the {\tt satmask} task (Richard Hook, private
communication) was used to generate a mask for each trail. This undistorted mask
was then transformed (using {\tt blot}) into the input distorted space of the
{\tt flt} images, assigned the unique flag value of 16384, and summed with the
existing {\tt flt} data quality array, e.g.{\tt
{j9l051gnq\_flt.fits{[}DQ,1{]}}}. Any such masking creates even more areas of
reduced signal-to-noise, and the affected areas are reflected in the weight
maps.

\section{Image combination}

The IRAF/STSDAS {\tt MultiDrizzle} task (Koekemoer et al., 2002) was used to
combine the mosaics for each filter. {\tt MultiDrizzle} is a PyRAF script which
performs, on a list of input {\tt flt} images: bad pixel identification, sky
subtraction, rejection of cosmic rays and other artifacts (as described above),
and a final {\tt drizzle} combination (Fruchter et al., 2002) with the cosmic
ray masks, into a clean image. {\tt MultiDrizzle} also applies the latest
filter-specific geometric distortion corrections to each image, as specified in
the IDCTAB reference tables. These mosaics were processed in the same
environment as the HST calibration pipeline (SunFire/smalls). See:

\url{http://www.stsci.edu/hst/acs/analysis/multidrizzle/}

Although the rejection of cosmic rays and other undesireable artifacts is
embedded in the {\tt MultiDrizzle} processing, the following is an overview of
how it is accomplished. A median image is constructed from the registered and
undistorted single-drizzled ({\tt \_single\_sci.fits}) images. This median image
-- or the appropriate sections of it -- are blotted back to the distorted space
of the input {\tt flt} images, where it can be used to identify cosmic rays. The
{\tt dither} package tasks of {\tt deriv} and {\tt driz\_cr} are used to compare
this blotted image and its derivative image with the original input {\tt flt}
file, and generate a cosmic ray mask. Finally, all the {\tt flt} images,
together with their newly created cosmic ray masks, are drizzled onto a single
output mosaic, which has units of  electrons per second in each pixel.

There are many {\tt MultiDrizzle} parameters which can be adjusted. The pipeline
uses carefully pre-defined sets of parameters  (defined in the MDRIZTAB table
identified in image headers), but in the standalone environment, optimal
combination parameters for a specific dataset can be found through some
trial-and-error iterations. Here we describe our key parameter settings,
especially any non-default choices.  The full set of drizzling parameters are
available in the README file on the MAST M82 webpage.

We set the center of the output mosaic to be the coordinates for M82 given by
the NASA Extragalactic Database  (in decimal degrees {\tt ra = 148.967580, dec =
69.679704}),  and rotate the output with north up, and east left ({\tt
final\_rot=0.0}). We set the output dimensions as 12288x12288 pixels, which  truncates a small number of pixels in the corners of
the rotated mosaic (Figure 1). We provide a shiftfile to refine the mosaic
registration as described in the previous section.
Since the galaxy fills most of the mosaic, sky subtraction was turned off 
({\tt skysub=no}). We set {\tt crbit=0} so that our rejections do not get flagged in the data quality arrays of the input {\tt flt} images. Note that the
weight maps reflect the cosmic rays and artifacts rejected by 
{\tt MultiDrizzle}. The cosmic ray rejection thresholds 
({\tt driz\_cr\_snr = 4.0 3.5}) were raised a bit to avoid rejecting the cores
of bright objects.   Since these mosaics are so large, we did not build them
into multi-extension  FITS files ({\tt build=no}), i.e. the science array and
weight maps are separate files, and we chose not to produce context images. Wet
set {\tt bits=4192} to ignore the flags for the warm pixels (flag=32) and CTE
tails  of hot pixels (flag=64), which are flagged in the dark reference image,
and any rejections flagged by pipeline processing  (flag=4096).
We use the default square drizzle kernel ({\tt final\_kernel=square}), which
performs well in most cases. Some correlated noise is evident as a faint
Moire-like pattern in the weight maps, which is also visible in the science data
-- especially in low signal-to-noise areas. Although the lanczos3 and gaussian
kernels can suppress this correlated noise better, they may also enhance
unrejected artifacts.

The cosmic ray masks generated by {\tt MultiDrizzle} are used as input to the
final drizzle combination of all the images. The {\tt drizzle} task (Fruchter
and Hook, 2002) performs a weighted sum of the input images, and allows input
pixels to be shrunk (we set {\tt pixfrac=0.9}) before being mapped onto the
output plane. The output pixel scale ({\tt final\_scale}) can also be different
than the input (detector) pixel scale -- a smaller pixel scale can allow more
spatial information to be recovered if the data are well-dithered. Since our
2-point sub-pixel dither pattern does not sample optimally in both dimensions,
even in conjuntion with the gap dither, we retain the input WFC pixel scale
(0.05 arcsec/pixel), rather than drizzle to a finer output scale.

{\tt MultiDrizzle} also produces exposure weight maps, which indicate the
background and instrumental noise for each pixel in the science data.  In
general, we acheived a relatively uniform exposure weight across the entire
mosaic. But the total exposure time can vary significantly from pixel-to-pixel,
mainly due to overlaps between adjacent tiles, interchip gaps, and all the bad
pixels which are rejected. The weight maps were visually inspected to ensure
that rejected artifacts have appropriately lower weight -- and equally
importantly, that real objects (e.g. the cores of stars, or of the core of M82)
were not rejected. The only irregular rejection was related to long diffraction
spikes and/or bleeding (from saturation) around bright foreground stars, which
is to be expected. Small amounts of residual artifacts or cosmic ray
contamination may have survived rejection -- most likely in areas of  low
weight, such as the gap overlaps.

The photometric fidelity of the {\tt MultiDrizzle} code is reliable to a high
degree of accuracy, since the underlying algorithm for drizzle is designed to be
flux-conserving (see Fruchter \& Hook, 2002, and Koekemoer et al., 2002). The
{\tt MultiDrizzle} team has verified this in practice by using a suite of test
datasets to compare photometry from different exposures of the same objects, and
verifying that these provide good agreement to better than the ACS flatfield and
photometric calibration accuracy (1-2\%) for bright sources, and Poisson noise
for fainter sources. But these M82 mosaics were not tested for photometric
accuracy, and were only spot-checked for PSF stability across the field.

\section{Data products and filenaming convention}

The drizzled mosaics described in this paper are available as
High-Level Science Products (HLSP) via the Multimission Archive at STScI (MAST):

\url{http://archive.stsci.edu/prepds/m82/}

The filenames of our drizzled science mosaics ({\tt \_drz\_sci.fits}) and their
corresponding exposure weight maps ({\tt \_drz\_weight.fits}), are listed in
Table 1. This filenaming convention allows these HLSP files to be queryable via
MAST archive searches. The filenames indicate the filter (b,v,i,h), and the
pixel scale of 0.05 arcsec (s05). These files have dimensions of 12288 x 12288
pixels (10.24 x 10.24 arcminutes) and are 604 MB each (uncompressed). Lower
resolution (1/4) block-averaged FITS versions of each mosaic were also produced
using the IRAF {\tt blkavg} task, mainly for quick-look downloading and viewing,
or for educational purposes (not intended for scientific analysis). These have a
scale of 0.20 arcsec per pixel (s20 in the filenames), dimensions of 3072 x 3072
pixels, and are 38 MB each, or 1/16 the filesize of the full-resolution mosaics
above. Much smaller preview GIF images of these mosaics are also available on
the MAST website. Note that the FITS data are all oriented with north up, and
east to the left (as in Figure 1), whereas the press release color composite
image (Figure 3) is at a position angle of 130 degrees. The color composite
image and related information are available from the associated STScI and Hubble
Heritage press releases:

\url{http://hubblesite.org/newscenter/newsdesk/archive/releases/2006/14/}

\url{http://heritage.stsci.edu/2006/14/}

\section{Acknowledgments}

Thanks to STScI Director Matt Mountain for granting 24 orbits of Director's
Discretionary observing time to the Hubble Heritage Team to conduct this
program. Thanks also to Warren Hack, Chris Hanley, Richard Hook, Marco
Sirianni, Anton Koekemoer, Inga Kamp, Karen Levay, Faith Abney, and Randy
Thompson for their contributions to this project.

If you utilize these prepared mosaics for further scientific analysis, please 
acknowledge and/or reference this paper.

\clearpage 

\begin{table}
\begin{center}
\caption{Filters, total exposure times, and mosaic filenames.\label{tbl-1}}
\begin{tabular}{crrrr}
\tableline\tableline
Filter & Exposure Time & Science Mosaic & Weight Map \\
\tableline
F435W (B) & 1600s & h\_m82\_b\_s05\_drz\_sci.fits & h\_m82\_b\_s05\_drz\_weight.fits \\
F555W (V) & 1360s & h\_m82\_v\_s05\_drz\_sci.fits & h\_m82\_v\_s05\_drz\_weight.fits \\
F814W (I) & 1360s & h\_m82\_i\_s05\_drz\_sci.fits & h\_m82\_i\_s05\_drz\_weight.fits  \\
F658N (H\( \alpha  \)) & 3320s & h\_m82\_h\_s05\_drz\_sci.fits & h\_m82\_h\_s05\_drz\_weight.fits \\
\tableline
\end{tabular}
\tablecomments{Via MAST, low resolution FITS and GIF versions of the files above are also available.}
\end{center}
\end{table}

\clearpage

\begin{figure}
\includegraphics[scale=1.00,angle=0]{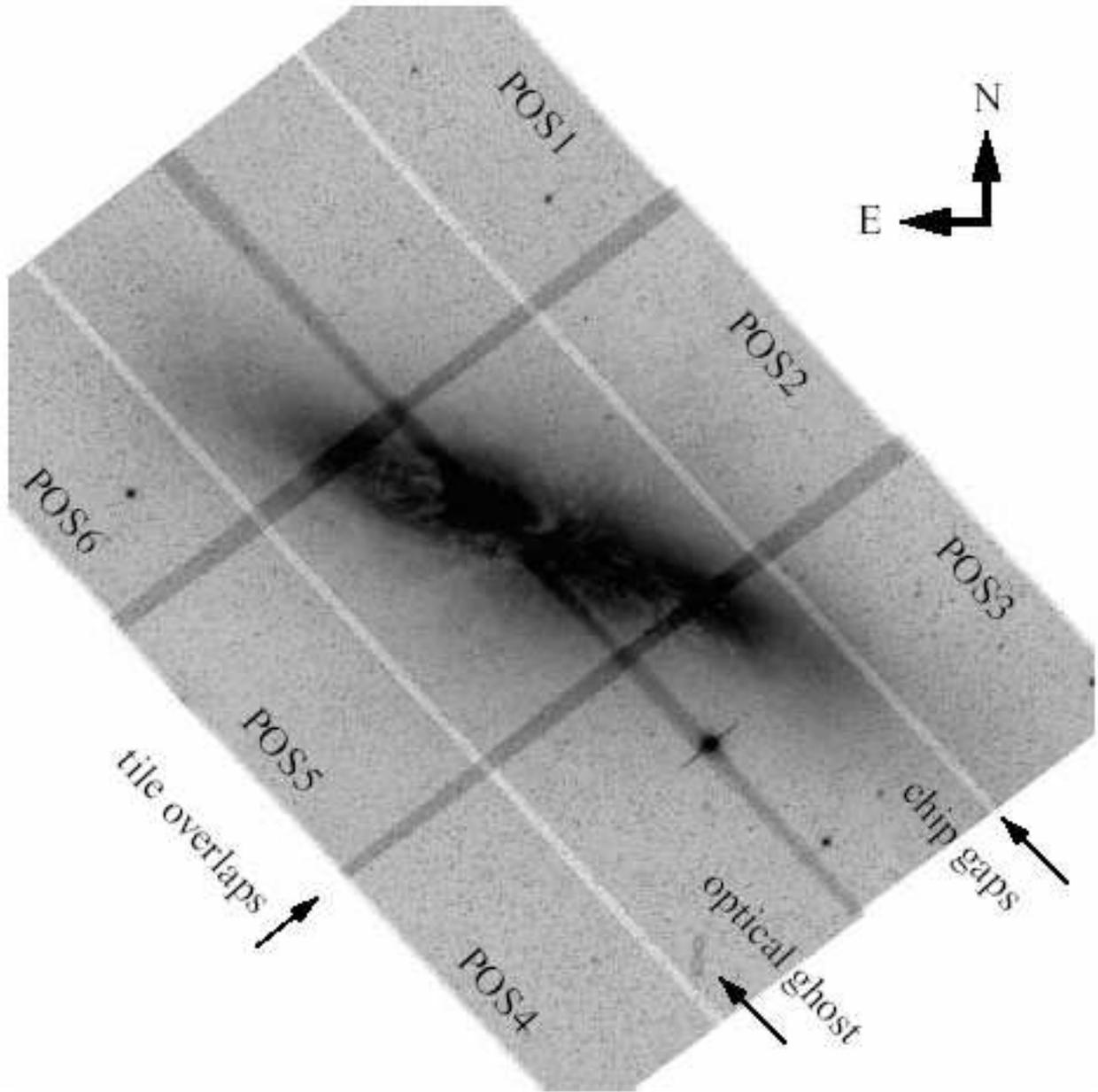}
\caption{Raw B-band (F435W) mosaic (intertiles only) illustrating the 
orientation and some key features of the 3x2 mosaic data. The mosaic files have
dimensions of 12288 x 12288 pixels (10.24 x 10.24 arcminutes). With north up
and east to the left, the data is truncated slightly in each corner.
\label{fig1}}
\end{figure}

\clearpage

\begin{figure}
\includegraphics[scale=1.50,angle=0]{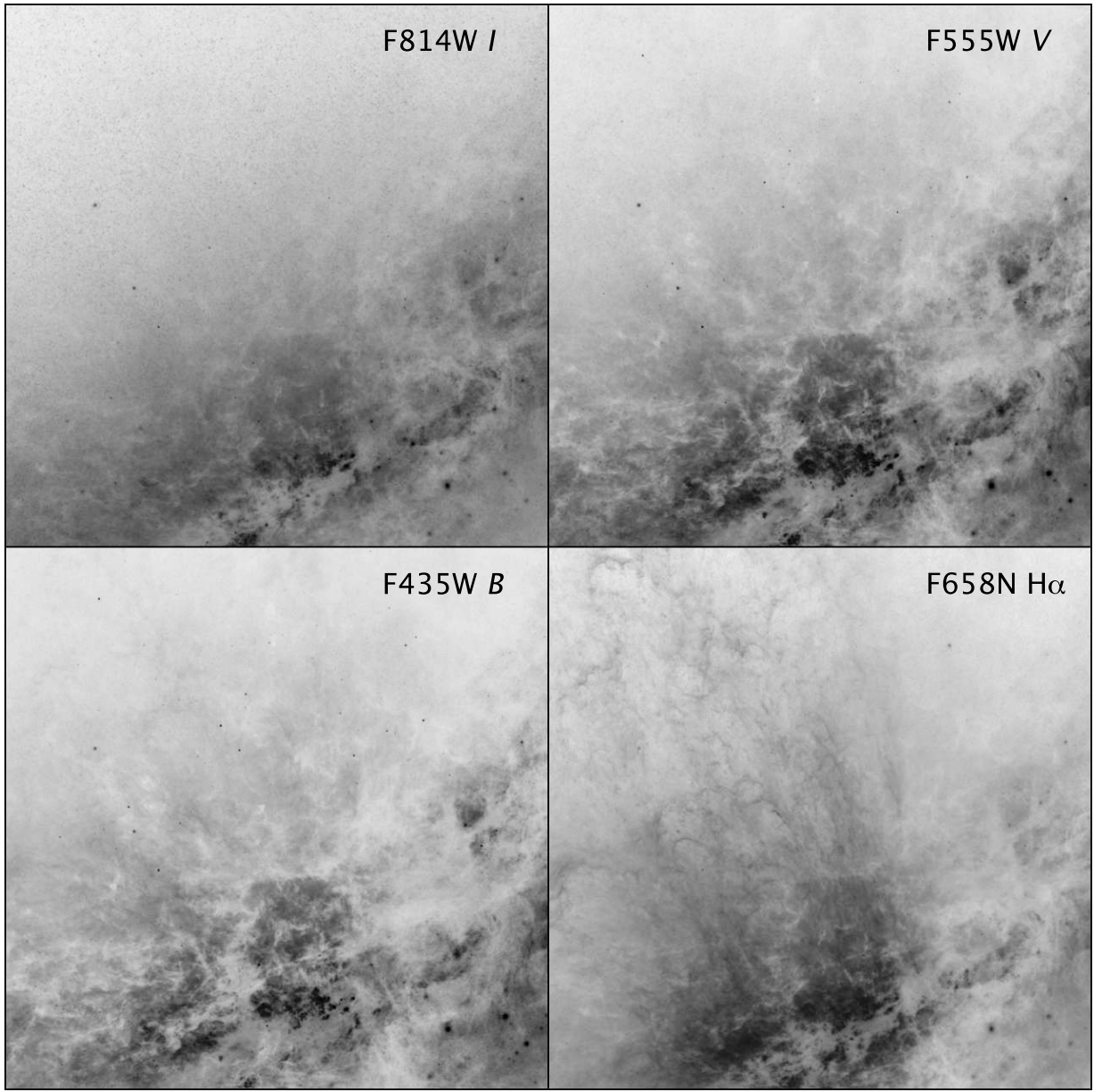}
\caption{Close-up of M82 galactic features in all four filters. North is not up
here.
\label{fig2}}
\end{figure}

\clearpage

\begin{figure}
\includegraphics[scale=0.50,angle=-90]{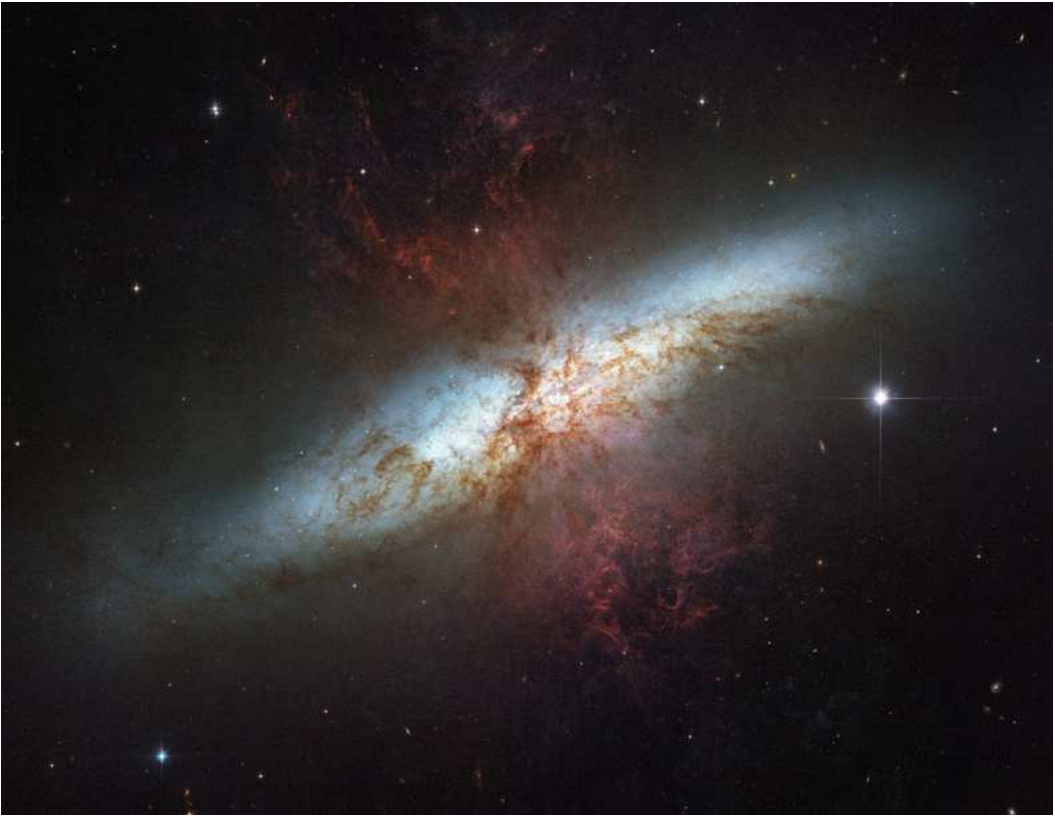}
\caption{Color composite mosaic of M82 made from all four filters. Note this
press release image is at a position angle of 130 degrees (north is not up
here).
\label{fig3}}
\end{figure}

\end{document}